\documentclass[11pt]{article}
\usepackage[margin=1in]{geometry}
\usepackage{amsmath, amssymb, amsthm}
\usepackage{booktabs}
\usepackage{hyperref}
\hypersetup{hidelinks}
\usepackage{enumitem}
\usepackage{setspace}
\newtheorem{assumption}{Assumption}

\DeclareMathOperator{\Cov}{Cov}
\DeclareMathOperator{\Var}{Var}
\DeclareMathOperator{\argmin}{arg\,min}
\DeclareMathOperator{\essinf}{ess\,inf}
\DeclareMathOperator{\esssup}{ess\,sup}

\title{Historical Developments in Probability Measures for Asset Pricing: From State Prices to Modern Pricing Kernels}
\author{Zhang Chen, Chen Kay}
\date{2026}

\begin{document}
\maketitle

\begin{abstract}
This review summarizes the historical development of probability measures in asset pricing, from early mathematical finance and state price theory to risk neutral valuation, martingale measures, forward measures, stochastic discount factors, incomplete market measure selection, benchmark pricing, robust and nonlinear pricing, and modern data driven probability transformations. The central theme is that asset pricing is not merely an exercise in estimating physical probabilities. Instead, pricing theory constructs, transforms, or selects probability measures so that market prices can be represented as expectations after discounting, numeraire normalization, marginal utility weighting, entropy penalization, calibration, or information conditioning. The paper gives special emphasis to the most influential publications that changed the field: Bachelier's probabilistic model of speculation, Arrow and Debreu's state contingent claims, Markowitz and the CAPM literature, Black Scholes and Merton's option pricing theory, Harrison Kreps and Harrison Pliska's martingale formalization, Delbaen and Schachermayer's general fundamental theorem, Breeden and Litzenberger's implied state price densities, Geman, El Karoui, and Rochet's change of numeraire, Hansen and Jagannathan's stochastic discount factor restrictions, Cochrane's SDF synthesis, and the modern empirical and machine learning literature on learned pricing kernels. Text, attention, and sentiment based probability transformations are treated as recent information-adjusted forecasting extensions that sit alongside, rather than above, the established martingale, numeraire, SDF, and incomplete market frameworks. The paper also states common semimartingale market assumptions and collects key mathematical formulas for state prices, stochastic discount factors, Radon Nikodym densities, Girsanov changes of measure, Black Scholes Merton valuation, forward measures, implied risk neutral densities, incomplete market selection rules, coherent risk measures, benchmark pricing, learned SDFs, and information adjusted forecasting. It concludes that probability measures in asset pricing have evolved from static state price ratios into a broad language for valuation, hedging, risk premia, ambiguity, information, and empirical identification.
\end{abstract}


\section{Introduction}

The history of asset pricing can be read as a history of probability measures. In elementary forecasting, a probability measure describes the likelihood of future states. In asset pricing, however, the relevant measure is usually not only the physical measure observed from historical frequencies. Prices also encode discounting, risk preferences, marginal utility, constraints, liquidity, market incompleteness, and information. For this reason, modern asset pricing repeatedly replaces, tilts, or augments the physical probability measure, usually denoted by $\mathbb{P}$, with another measure or valuation object under which prices become conditional expectations.

The simplest statement is familiar. If $H_T$ is a payoff at maturity $T$, then in a complete arbitrage free market its time zero price can often be represented as
\[
V_0=\mathbb{E}^{\mathbb{Q}}\left[\exp\left(-\int_0^T r_s ds\right)H_T\right],
\]
where $\mathbb{Q}$ is a risk neutral or equivalent martingale measure. Yet this compact equation hides more than a century of conceptual development. It requires a state space, traded claims, a discounting convention, a relation between prices and probability weights, and a theorem ensuring that no arbitrage is equivalent to the existence of a suitable probability measure. It also requires a selection principle when the market is incomplete, since there may be many such measures.

The purpose of this review is to summarize the main historical developments in probability measures for asset pricing. The paper covers early mathematical foundations, the Arrow Debreu state price view, the Black Scholes Merton risk neutral transformation, martingale pricing theory, change of numeraire, term structure measures, incomplete market measures, stochastic discount factors, benchmark and real world pricing, robust and nonlinear approaches, and recent data driven reweighting methods. The final part of the review turns to data rich environments in which text, attention, and sentiment enter the information set used for forecasting or for estimating a pricing kernel. These recent approaches are presented as continuations of the broader measure change tradition: they modify the distribution used for prediction or conditioning, while no arbitrage valuation remains governed by martingale measures, numeraires, and stochastic discount factors.

Several organizing distinctions are useful. The physical measure $\mathbb{P}$ describes statistical dynamics. The risk neutral measure $\mathbb{Q}$ removes risk premia from discounted asset dynamics when valuation can be done by replication. A forward measure uses a zero coupon bond or other numeraire so that forward prices become martingales. A pricing kernel or stochastic discount factor represents valuation under the physical measure by multiplying payoffs by a random marginal valuation term. In incomplete markets, equivalent martingale measures are no longer unique, so one may choose a measure through hedging error, entropy, variance, utility, calibration, robustness, or benchmark arguments. In data rich settings, text, attention, and sentiment can enter either as conditioning variables in a learned stochastic discount factor or as information adjusted probability tilts for forecasting.

\section{Landmark Publications and Historical Emphasis}

This review emphasizes publications that changed how probability was used in asset pricing, not merely papers that introduced additional notation. A publication is treated as historically central when it did at least one of the following: introduced a new valuation object, established a measure existence or uniqueness theorem, connected pricing measures to economic equilibrium, made a probability measure empirically recoverable from market prices, or changed the empirical estimation of the stochastic discount factor. Table \ref{tab:landmarks} gives a compact guide to the most important milestones.

\begin{table}[!htbp]
\centering
\small
\renewcommand{\arraystretch}{1.14}
\begin{tabular}{p{0.19\textwidth}p{0.36\textwidth}p{0.36\textwidth}}
\toprule
Period & Representative landmark publications & Measure-theoretic contribution \\
\midrule
1900--1960s & Bachelier; Arrow; Debreu; Samuelson \cite{Bachelier1900,Arrow1953,Debreu1959,Samuelson1965} & Stochastic price dynamics and state contingent claims; normalized state prices anticipate risk neutral probabilities. \\
1950s--1970s & Markowitz; Sharpe; Lintner; Mossin; Ross \cite{Markowitz1952,Sharpe1964,Lintner1965,Mossin1966,Ross1976} & Physical expected returns are linked to covariance and factor risk; pricing kernels later unify these equilibrium restrictions. \\
1973--1979 & Black and Scholes; Merton; Cox, Ross, and Rubinstein \cite{BlackScholes1973,Merton1973,CoxRossRubinstein1979} & Dynamic replication removes physical drift and makes risk neutral probabilities operational in option pricing. \\
1978--1995 & Breeden and Litzenberger; Heston; Duffie, Pan, and Singleton \cite{BreedenLitzenberger1978,Heston1993,DuffiePanSingleton2000} & Option prices reveal risk neutral densities; stochastic volatility, jumps, and affine models enrich the pricing measure. \\
1979--1998 & Harrison and Kreps; Harrison and Pliska; Delbaen and Schachermayer \cite{HarrisonKreps1979,HarrisonPliska1981,DelbaenSchachermayer1994,DelbaenSchachermayer1998} & Equivalent martingale measures become the rigorous expression of no arbitrage; completeness corresponds to uniqueness. \\
1980s--1990s & Vasicek; Cox, Ingersoll, and Ross; Heath, Jarrow, and Morton; Geman, El Karoui, and Rochet \cite{Vasicek1977,CoxIngersollRoss1985,HeathJarrowMorton1992,GemanElKarouiRochet1995} & Term structure models and change of numeraire show that numeraires generate natural forward measures. \\
1990s--2000s & Hansen and Jagannathan; Lucas; Cochrane; Fama and French \cite{HansenJagannathan1991,Lucas1978,Cochrane2005,FamaFrench1993} & Stochastic discount factors become the common language for equilibrium, empirical factor models, and no arbitrage pricing. \\
1990s--2010s & Follmer and Schweizer; Frittelli; Artzner et al.; Hansen and Sargent \cite{FollmerSchweizer1991,Frittelli2000,ArtznerDelbaenEberHeath1999,HansenSargent2008} & Incompleteness, entropy, utility, coherent risk, and ambiguity transform measure choice into an optimization or robustness problem. \\
2010s--2026 & Gu, Kelly, and Xiu; Chen, Pelger, and Zhu; Kelly et al.; Korsaye, Quaini, and Trojani; Wang, Cheng, and Wang; Avramov and He \cite{GuKellyXiu2020,ChenPelgerZhu2024,KellyKuznetsovMalamudXu2025,KorsayeQuainiTrojani2025,WangChengWang2025,AvramovHe2026} & High dimensional data estimate pricing kernels and learned measure changes; text, news, and attention become conditioning information for empirical SDFs and forecasting laws. \\
\bottomrule
\end{tabular}
\caption{Landmark publications emphasized in the review. The table is selective: it highlights works that changed the role of probability measures, state prices, or stochastic discount factors in asset pricing.}
\label{tab:landmarks}
\end{table}

The remaining sections follow this hierarchy. Foundational and widely adopted results receive the most space. More recent information adjusted methods are included to show continuity with measure transformation, but they are presented after the classical martingale, numeraire, incomplete market, and stochastic discount factor literatures.

\clearpage
\section{Mathematical Framework, Assumptions, and Notation}

Because this review follows probability measures across many asset pricing models, it is useful to state the common mathematical environment before turning to the chronology. The notation below is deliberately broad: it includes finite state Arrow Debreu economies, diffusion models, semimartingale markets, incomplete markets, and information adjusted forecasting measures.

\begin{assumption}[Filtered probability space]
There is a finite horizon $T>0$ and a filtered probability space $(\Omega,\mathcal{F},(\mathcal{F}_t)_{0\leq t\leq T},\mathbb{P})$ satisfying the usual conditions. The measure $\mathbb{P}$ is called the physical, historical, or statistical measure. It describes the data generating law of cash flows, returns, signals, and news. Conditional expectations are taken with respect to the available information $\mathcal{F}_t$.
\end{assumption}

\begin{assumption}[Traded assets and discounting]
The economy contains a strictly positive savings account
\[
B_t=\exp\left(\int_0^t r_s\,ds\right),
\]
and $d$ risky assets $S^1,\ldots,S^d$. The discounted price vector is
\[
\widetilde S_t=\frac{S_t}{B_t}=(\widetilde S^1_t,\ldots,\widetilde S^d_t),
\qquad \widetilde S^i_t:=B_t^{-1}S^i_t.
\]
In continuous time, $S$ is assumed to be a semimartingale, which is the natural class of processes for stochastic integration and self financing trading.
\end{assumption}

\begin{assumption}[Self financing and admissibility]
A predictable portfolio process $\theta=(\theta^1,\ldots,\theta^d)$ is self financing if the discounted wealth satisfies
\[
\widetilde X_t=x+\int_0^t \theta_u\,d\widetilde S_u.
\]
Admissible strategies are restricted so that wealth is bounded from below, or satisfies an equivalent integrability condition, excluding doubling strategies that would generate artificial arbitrage.
\end{assumption}

\begin{assumption}[No arbitrage]
The benchmark absence of arbitrage condition in modern continuous time theory is no free lunch with vanishing risk (NFLVR). Under the fundamental theorem of asset pricing, NFLVR is equivalent, under suitable technical conditions, to the existence of at least one equivalent local martingale measure for discounted prices \cite{DelbaenSchachermayer1994,DelbaenSchachermayer1998}.
\end{assumption}

\begin{assumption}[Payoffs and integrability]
A contingent claim is an $\mathcal{F}_T$ measurable random variable $H$. Pricing formulas below require integrability under the relevant measure, for example
\[
\mathbb{E}^{\mathbb{Q}}\left[B_T^{-1}|H|\right]<\infty
\]
for risk neutral valuation, or
\[
\mathbb{E}^{\mathbb{P}}\left[|M_T H|\right]<\infty
\]
when a stochastic discount factor $M$ is used.
\end{assumption}

\subsection{Equivalent martingale measures and the pricing formula}

The central object of arbitrage based pricing is the set
\[
\mathcal{M}_e=\left\{\mathbb{Q}\sim\mathbb{P}: \widetilde S \text{ is a } \mathbb{Q}\text{-local martingale}\right\}.
\]
If $\mathcal{M}_e\neq\varnothing$, then the market is arbitrage free in the NFLVR sense. If $\mathcal{M}_e$ is a singleton, the market is complete in the usual diffusion setting; if it contains many elements, the market is incomplete and additional criteria are needed to select a pricing measure.

For a replicable payoff $H$, the arbitrage free price process is
\begin{equation}
V_t=B_t\,\mathbb{E}^{\mathbb{Q}}\left[B_T^{-1}H\mid\mathcal{F}_t\right],
\qquad 0\leq t\leq T,
\label{eq:riskneutralpricing}
\end{equation}
for every $\mathbb{Q}\in\mathcal{M}_e$. In an incomplete market, nonreplicable claims often have no unique price. A common no arbitrage interval is
\begin{equation}
\underline V_t=\essinf_{\mathbb{Q}\in\mathcal{M}_e}B_t\mathbb{E}^{\mathbb{Q}}\left[B_T^{-1}H\mid\mathcal{F}_t\right],
\qquad
\overline V_t=\esssup_{\mathbb{Q}\in\mathcal{M}_e}B_t\mathbb{E}^{\mathbb{Q}}\left[B_T^{-1}H\mid\mathcal{F}_t\right].
\label{eq:noarbitrageinterval}
\end{equation}
Measure selection rules such as minimal martingale, variance optimal, minimal entropy, and utility indifference measures narrow this interval by adding hedging or preference criteria.

\subsection{Radon Nikodym densities and Bayes pricing}

If $\mathbb{Q}\sim\mathbb{P}$ on $\mathcal{F}_T$, its density process is
\begin{equation}
Z_t=\mathbb{E}^{\mathbb{P}}\left[\frac{d\mathbb{Q}}{d\mathbb{P}}\bigg|\mathcal{F}_t\right],
\qquad Z_0=1,\qquad Z_t>0.
\label{eq:densityprocess}
\end{equation}
For any integrable $Y$, Bayes' rule gives
\begin{equation}
\mathbb{E}^{\mathbb{Q}}[Y\mid\mathcal{F}_t]
=\frac{1}{Z_t}\mathbb{E}^{\mathbb{P}}[Z_TY\mid\mathcal{F}_t].
\label{eq:bayes}
\end{equation}
Thus risk neutral pricing can be written under the physical measure as
\begin{equation}
V_t=\frac{B_t}{Z_t}\,\mathbb{E}^{\mathbb{P}}\left[Z_TB_T^{-1}H\mid\mathcal{F}_t\right].
\label{eq:ppricingdensity}
\end{equation}
The product
\begin{equation}
M_t=\frac{Z_t}{B_t}
\label{eq:sdfdensity}
\end{equation}
is a stochastic discount factor or state price density. Equation \eqref{eq:ppricingdensity} becomes
\begin{equation}
V_t=\frac{1}{M_t}\mathbb{E}^{\mathbb{P}}[M_T H\mid\mathcal{F}_t].
\label{eq:sdfpricing}
\end{equation}
This identity is the bridge between no arbitrage derivative pricing and equilibrium asset pricing.

\subsection{Girsanov theorem and the market price of risk}

In a Brownian diffusion model under $\mathbb{P}$,
\begin{equation}
\frac{dS_t}{S_t}=\mu_t\,dt+\sigma_t\,dW_t^{\mathbb{P}},
\label{eq:pmeasurestock}
\end{equation}
where $\mu_t$ is a vector of expected returns and $\sigma_t$ is a volatility matrix. If there exists a progressively measurable process $\lambda_t$ such that
\begin{equation}
\mu_t-r_t\mathbf{1}=\sigma_t\lambda_t,
\label{eq:marketpriceofrisk}
\end{equation}
then $\lambda_t$ is a market price of risk. Under Novikov type integrability conditions, the stochastic exponential
\begin{equation}
Z_T^{\lambda}=\mathcal{E}\left(-\int_0^T \lambda_u^\top dW_u^{\mathbb{P}}\right)
=\exp\left(-\int_0^T \lambda_u^\top dW_u^{\mathbb{P}}-\frac12\int_0^T\|\lambda_u\|^2du\right)
\label{eq:girsanovdensity}
\end{equation}
defines a measure $\mathbb{Q}$ by $d\mathbb{Q}=Z_T^\lambda d\mathbb{P}$. Girsanov's theorem states that
\begin{equation}
dW_t^{\mathbb{Q}}=dW_t^{\mathbb{P}}+\lambda_t\,dt
\label{eq:girsanovbrownian}
\end{equation}
is Brownian under $\mathbb{Q}$, and therefore
\begin{equation}
\frac{dS_t}{S_t}=r_t\,dt+\sigma_t\,dW_t^{\mathbb{Q}}.
\label{eq:qmeasurestock}
\end{equation}
This is the continuous time mathematical expression of the Black Scholes Merton insight: under the pricing measure, expected returns are replaced by the short rate while volatility and payoff exposure remain.

\subsection{Finite state state prices}

In a one period finite state economy with states $\omega_1,\ldots,\omega_n$, let $\pi_i$ be the price of an Arrow claim paying one unit in state $i$. For a payoff $X=(X_1,\ldots,X_n)$,
\begin{equation}
p(X)=\sum_{i=1}^n \pi_i X_i.
\label{eq:stateprice}
\end{equation}
If the risk free gross return is $R_f$, the risk neutral probabilities are
\begin{equation}
q_i=R_f\pi_i,
\qquad q_i>0,
\qquad \sum_{i=1}^n q_i=1,
\label{eq:finiteq}
\end{equation}
and the price is
\begin{equation}
p(X)=\frac{1}{R_f}\mathbb{E}^{\mathbb{Q}}[X].
\label{eq:finitepricing}
\end{equation}
The formula shows that risk neutral probabilities are normalized state prices, not necessarily empirical frequencies.

\subsection{One period stochastic discount factors}

Let $R_i$ be the gross return on asset $i$. A one period stochastic discount factor $m$ satisfies
\begin{equation}
\mathbb{E}^{\mathbb{P}}[mR_i]=1,
\qquad i=1,\ldots,d.
\label{eq:onedaysdf}
\end{equation}
Equivalently, for any payoff $X$, $p(X)=\mathbb{E}^{\mathbb{P}}[mX]$. This equation is the foundation of empirical asset pricing. It implies, for a risk free return $R_f$, that $\mathbb{E}^{\mathbb{P}}[m]=1/R_f$, and for any risky asset,
\begin{equation}
\mathbb{E}^{\mathbb{P}}[R_i]-R_f
=-R_f\,\Cov^{\mathbb{P}}(m,R_i).
\label{eq:riskpremiumcov}
\end{equation}
Risk premia are therefore covariance prices: assets that covary negatively with marginal value or the pricing kernel must offer high expected returns. The Hansen Jagannathan bound formalizes this implication as a lower bound on the volatility of admissible discount factors \cite{HansenJagannathan1991}.

\subsection{A map of measure terminology}

The following identities summarize the main objects used in the remainder of the review:
\begin{align}
\text{Physical law:} \quad &\mathbb{P}, \label{eq:phys}\\
\text{Pricing law:} \quad &\mathbb{Q}\sim\mathbb{P},\quad \widetilde S \text{ is a }\mathbb{Q}\text{-martingale}, \label{eq:pricinglaw}\\
\text{Density process:} \quad &Z_t=\frac{d\mathbb{Q}}{d\mathbb{P}}\bigg|_{\mathcal F_t}, \label{eq:density}\\
\text{State price density:} \quad &M_t=Z_t/B_t, \label{eq:statepricedensity}\\
\text{Forward measure:} \quad &\frac{d\mathbb{Q}^{N}}{d\mathbb{Q}}\bigg|_{\mathcal F_T}=\frac{B_0N_T}{N_0B_T}, \label{eq:forwardmeasure}\\
\text{Robust family:} \quad &\mathcal{Q}\subseteq\{\mathbb{Q}:\mathbb{Q}\ll\mathbb{P}\}, \label{eq:robustfamily}\\
\text{Information adjusted law:} \quad &\frac{d\mathbb{P}^{a}}{d\mathbb{P}}=Z^a,\quad Z^a>0,\quad \mathbb{E}^{\mathbb{P}}[Z^a]=1. \label{eq:infoadjustedlaw}
\end{align}
The historical sections below can be viewed as successive answers to three questions: which measure is economically meaningful, how is it constructed, and what assumptions guarantee that it prices assets or improves forecasts?

\section{From Bachelier to State Prices}

The mathematical origin of asset pricing is often traced to Bachelier's 1900 thesis on speculation, which introduced Brownian motion as a model for price fluctuations before Einstein's physical work on Brownian motion \cite{Bachelier1900}. Bachelier did not formulate a risk neutral measure in the modern sense, but his work introduced the crucial idea that security prices could be modeled probabilistically. Later, Samuelson developed geometric Brownian motion and placed stochastic processes at the center of financial economics \cite{Samuelson1965}. These early models operated mostly under a physical or statistical measure, with expected returns treated as primitives rather than objects to be transformed away by arbitrage.

The first decisive step toward pricing measures came from state contingent claims. Arrow introduced securities that pay one unit in one state and zero in all others \cite{Arrow1953}. Debreu then embedded contingent commodities in general equilibrium theory \cite{Debreu1959}. In a finite state economy with states $\omega_1,\ldots,\omega_n$, an Arrow security has payoff $1$ in state $i$ and $0$ elsewhere. If its time zero price is $\pi_i$, then a claim $X=(X_1,\ldots,X_n)$ has price
\[
p(X)=\sum_{i=1}^n \pi_i X_i.
\]
If there is a risk free asset with gross return $R$, then one may define risk neutral probabilities by
\[
q_i=R\pi_i,
\qquad \sum_{i=1}^n q_i=1,
\]
provided the state prices are positive and sum to $1/R$. Then
\[
p(X)=\frac{1}{R}\sum_{i=1}^n q_i X_i.
\]
Thus risk neutral probabilities can be interpreted as normalized state prices. They are not necessarily true probabilities in the statistical sense, but they satisfy the axioms of probability and price all payoffs by discounted expectation. This state price view remains one of the cleanest ways to understand risk neutral measures.

The Arrow Debreu framework also clarified the relation between completeness and uniqueness. If all state claims can be replicated by traded securities, then the state price vector is unique. If the market is incomplete, many positive state price vectors may be consistent with observed prices. This finite state intuition anticipates the fundamental theorem of asset pricing and the multiplicity of equivalent martingale measures in incomplete continuous time markets.

\section{Mean Variance Equilibrium and the Price of Risk}

Before martingale pricing became standard, asset pricing theory developed through equilibrium and mean variance analysis. Markowitz introduced portfolio selection by optimizing mean and variance \cite{Markowitz1952}. Sharpe, Lintner, and Mossin developed the capital asset pricing model, in which expected excess returns are proportional to market beta \cite{Sharpe1964,Lintner1965,Mossin1966}. Ross introduced arbitrage pricing theory, which connected expected returns to factor exposures under weaker equilibrium assumptions \cite{Ross1976}. Later empirical factor models, especially the Fama French three factor model, made the factor view central to applied asset pricing and reinforced the idea that the pricing kernel can often be summarized by a small number of systematic risks \cite{FamaFrench1993}.

These models do not usually define a risk neutral measure explicitly. Nevertheless, they contain the economic content behind the change from physical probabilities to pricing probabilities. Under the physical measure, a risky asset satisfies
\[
\mathbb{E}^{\mathbb{P}}[R_i]-R_f=\beta_i\left(\mathbb{E}^{\mathbb{P}}[R_M]-R_f\right)
\]
in the CAPM. The expected return is not the risk free rate because investors require compensation for systematic risk. Risk neutral pricing can be understood as shifting from a probability measure under which expected returns include risk premia to a measure under which discounted prices have zero drift. The market price of risk is the bridge between the two.

The stochastic discount factor formulation later unified equilibrium and no arbitrage views. A payoff $X$ has price
\[
p(X)=\mathbb{E}^{\mathbb{P}}[mX],
\]
where $m$ is a positive stochastic discount factor. If $m$ is normalized by the risk free discount factor, it induces a risk neutral measure through a Radon Nikodym derivative. Thus the state price density, pricing kernel, marginal utility growth, and risk neutral density are different faces of the same object \cite{HansenJagannathan1991,Cochrane2005}.

\section{The Black Scholes Merton Revolution}

The modern risk neutral measure became central after the option pricing breakthrough of Black and Scholes and Merton \cite{BlackScholes1973,Merton1973}. In the Black Scholes model, the stock price under the physical measure follows
\[
dS_t=\mu S_tdt+\sigma S_tdW_t^{\mathbb{P}},
\]
while the money market account satisfies $dB_t=rB_tdt$. The option price can be derived by dynamic replication, leading to a partial differential equation in which the physical drift $\mu$ disappears. Equivalently, one introduces a probability measure $\mathbb{Q}$ under which
\[
dS_t=rS_tdt+\sigma S_tdW_t^{\mathbb{Q}}.
\]
The option price is then
\[
V_t=\mathbb{E}^{\mathbb{Q}}\left[e^{-r(T-t)}H(S_T)\mid\mathcal{F}_t\right].
\]
This is the risk neutral valuation formula.

The binomial model of Cox, Ross, and Rubinstein made the same idea transparent in discrete time \cite{CoxRossRubinstein1979}. In a one step tree with up return $u$, down return $d$, and gross risk free return $R$, the risk neutral probability is
\[
q=\frac{R-d}{u-d},
\qquad 1-q=\frac{u-R}{u-d},
\]
provided $d<R<u$. Option values are then discounted expectations under $(q,1-q)$. This construction is historically important because it shows that risk neutral probabilities are normalized no-arbitrage prices, not empirical frequencies.

The conceptual importance is that the option price does not depend on the expected stock return $\mu$. The investor's required return is replaced by replication. The probability measure used for valuation is not a subjective belief about future stock prices, but an arbitrage consistent probability measure implied by the traded stock and bond. The Black Scholes Merton result therefore transformed probability from a statistical input into a pricing device.

Girsanov's theorem provides the mathematical mechanism. If the market price of risk is
\[
\lambda=\frac{\mu-r}{\sigma},
\]
then the Radon Nikodym density process is
\[
Z_t=\exp\left(-\lambda W_t^{\mathbb{P}}-\frac{1}{2}\lambda^2t\right).
\]
Defining $d\mathbb{Q}/d\mathbb{P}$ on $\mathcal{F}_t$ by $Z_t$, the process
\[
W_t^{\mathbb{Q}}=W_t^{\mathbb{P}}+\lambda t
\]
is a Brownian motion under $\mathbb{Q}$. The drift adjustment changes expected returns while preserving volatility and equivalent null sets. This measure transformation became the technical engine of continuous time asset pricing \cite{Girsanov1960,LiptserShiryaev1977,KaratzasShreve1998}.

\section{Martingale Measures and the Fundamental Theorem}

The next historical milestone was the martingale formalization of arbitrage pricing. Harrison and Kreps introduced martingales and multiperiod security markets into mathematical finance \cite{HarrisonKreps1979}. Harrison and Pliska developed the continuous time martingale theory of stochastic integrals and trading strategies \cite{HarrisonPliska1981}. The central insight is that absence of arbitrage is equivalent, under suitable technical conditions, to the existence of a probability measure equivalent to $\mathbb{P}$ under which discounted asset prices are martingales.

Let $S_t$ be the vector of risky asset prices and $B_t$ the money market account. A probability measure $\mathbb{Q}$ equivalent to $\mathbb{P}$ is an equivalent martingale measure if
\[
\frac{S_t}{B_t}
\]
is a martingale under $\mathbb{Q}$. Then every self financing wealth process discounted by $B_t$ is also a local martingale under $\mathbb{Q}$, ruling out arbitrage in an appropriate sense. Conversely, no arbitrage implies the existence of a separating linear functional, which can be represented as a positive state price density or martingale measure.

The fundamental theorem of asset pricing was later refined in increasingly general settings by Delbaen and Schachermayer, who introduced conditions such as no free lunch with vanishing risk \cite{DelbaenSchachermayer1994,DelbaenSchachermayer1998}. These developments solved technical problems that arise in infinite dimensional, continuous time settings, where simple no arbitrage can be too weak. The theorem gave rigorous meaning to the practical statement that arbitrage free pricing is equivalent to valuation under some martingale measure.

The second fundamental theorem connects completeness to uniqueness. In complete markets, every contingent claim can be replicated, and the equivalent martingale measure is unique. In incomplete markets, many equivalent martingale measures may exist, producing a range of no arbitrage prices. This distinction drives much of the later history of probability measures for asset pricing.

\section{Change of Numeraire and Forward Measures}

Once risk neutral valuation was established, researchers realized that the choice of numeraire also determines the convenient probability measure. Geman, El Karoui, and Rochet developed the change of numeraire technique, showing that if $N_t$ is a strictly positive traded numeraire, then asset prices divided by $N_t$ are martingales under an associated measure $\mathbb{Q}^N$ \cite{GemanElKarouiRochet1995}. If $V_t$ is the price of a payoff $H_T$, then
\[
\frac{V_t}{N_t}=\mathbb{E}^{\mathbb{Q}^N}\left[\frac{H_T}{N_T}\mid\mathcal{F}_t\right].
\]
This result made clear that there is not one privileged risk neutral measure. There is a family of numeraire associated pricing measures.

The most important application is the forward measure. If $P(t,T)$ is the price at time $t$ of a zero coupon bond maturing at $T$, then using $P(t,T)$ as numeraire defines the $T$ forward measure. Under this measure, forward prices for maturity $T$ are martingales. This greatly simplifies interest rate derivative pricing. Earlier short rate models such as Vasicek and Cox Ingersoll Ross specified a risk neutral short rate process and made bond prices conditional expectations of discount factors \cite{Vasicek1977,CoxIngersollRoss1985}. The Heath Jarrow Morton framework then modeled the whole forward rate curve and used drift restrictions to ensure arbitrage free dynamics under a chosen measure \cite{HeathJarrowMorton1992}. Brace, Gatarek, and Musiela developed the market model for LIBOR rates, whose natural measures are forward measures associated with bond numeraires \cite{BraceGatarekMusiela1997}. Jamshidian, Musiela and Rutkowski, and Bjork further developed the theory and applications of term structure measures \cite{Jamshidian1989,MusielaRutkowski2005,Bjork2009}.

The change of numeraire perspective is not just a technical trick. It shows that probability measures in asset pricing are relative to units of account. Cash, a zero coupon bond, a stock, a growth optimal portfolio, or a foreign money market account can each define a different measure. In foreign exchange, for example, domestic and foreign risk neutral measures differ by the exchange rate numeraire. This is a precursor to later ideas about benchmark pricing and numeraire invariance.

\section{Risk Neutral Densities and Implied Probability Measures}

Option markets allow the extraction of implied risk neutral distributions. Breeden and Litzenberger showed that the second derivative of a call price with respect to strike identifies the risk neutral density of the underlying terminal price \cite{BreedenLitzenberger1978}. If $C(K,T)$ is the price of a European call with strike $K$, then under regularity conditions
\[
\frac{\partial^2 C}{\partial K^2}=e^{-rT}f_{\mathbb{Q}}(K),
\]
where $f_{\mathbb{Q}}$ is the risk neutral density of $S_T$.

This result made the risk neutral measure observable, at least indirectly, from a cross section of option prices, and later nonparametric work by Ait-Sahalia and Lo reinforced the empirical recovery of state price densities \cite{AitSahaliaLo1998}. It also clarified the difference between historical and implied probabilities. The implied distribution contains risk premia and market pricing of tail states. Empirically, risk neutral densities extracted from options are often more negatively skewed and have heavier tails than physical densities, reflecting crash risk premia and demand for insurance \cite{Rubinstein1994,Jackwerth2000}. The volatility smile after the 1987 crash reinforced the view that a single lognormal risk neutral distribution was inadequate \cite{DermanKani1994,Dupire1994,Rubinstein1994}.

Dupire's local volatility model is especially important in the history of measures because it shows how the entire implied volatility surface can be represented by a diffusion under the risk neutral measure \cite{Dupire1994}. Rather than estimating physical drift, one calibrates a risk neutral diffusion to market option prices. Heston's stochastic volatility model similarly specified tractable risk neutral dynamics with stochastic variance \cite{Heston1993}. Later affine jump diffusion models and exponential Levy models extended the menu of risk neutral distributions used for pricing \cite{Bates1996,DuffiePanSingleton2000,ContTankov2004}.

\section{Incomplete Markets and Measure Selection}

In complete Black Scholes markets, the risk neutral measure is unique. Realistic markets are incomplete because of jumps, stochastic volatility, transaction costs, discrete trading, unhedgeable factors, default risk, liquidity risk, and nontraded state variables. In incomplete markets, equivalent martingale measures form a set. Pricing a claim requires either selecting one measure, specifying preferences, or accepting price intervals.

Several measure selection principles emerged. Follmer and Sondermann developed mean variance hedging and the minimal martingale measure, which changes only the martingale component linked to traded risk while leaving orthogonal unhedgeable risk unchanged \cite{FollmerSondermann1986,FollmerSchweizer1991}. The variance optimal martingale measure minimizes squared hedging error and is closely related to quadratic hedging \cite{Schweizer1995}. The minimal entropy martingale measure selects the martingale measure closest to the physical measure in relative entropy \cite{Frittelli2000,GranditsRheinlander2002}. Utility indifference pricing selects prices by comparing expected utility with and without the claim, leading to probability changes determined by investor preferences \cite{HodgesNeuberger1989,Davis1997,ManiaSchweizer2005}.

These approaches show that probability measures are not merely discovered, they are chosen according to an economic criterion. The choice reflects what one wants to preserve, minimize, or optimize. The minimal martingale measure preserves unhedgeable noise, the minimal entropy measure penalizes information distance from $\mathbb{P}$, and the utility based measure reflects marginal utility. Thus incomplete market asset pricing transformed the question from ``what is the risk neutral measure?'' to ``which pricing measure is appropriate for this hedging, preference, or calibration problem?''

Defaultable securities provide another example. Reduced form credit models often use a risk neutral default intensity, which differs from physical default probabilities because it incorporates credit risk premia \cite{JarrowTurnbull1995,DuffieSingleton1999}. Structural models begin with firm value dynamics and default boundaries, then transform to risk neutral valuation \cite{Merton1974}. The difference between physical and risk neutral default probabilities is central to credit spreads.

\section{Pricing Kernels, Stochastic Discount Factors, and Equilibrium Measures}

Parallel to martingale pricing, asset pricing economics developed the stochastic discount factor approach. Hansen and Jagannathan characterized restrictions on admissible discount factors using asset return data \cite{HansenJagannathan1991}. Cochrane presented the stochastic discount factor as the unifying object of modern asset pricing \cite{Cochrane2005}. The Euler equation
\[
1=\mathbb{E}^{\mathbb{P}}[m_{t+1}R_{t+1}]
\]
shows that $m_{t+1}$ prices all returns. If there is a risk free return $R_f$, then $\mathbb{E}^{\mathbb{P}}[m]=1/R_f$. A risk neutral measure can be defined by
\[
\frac{d\mathbb{Q}}{d\mathbb{P}}=R_f m.
\]
Thus the risk neutral measure is the physical measure tilted by the stochastic discount factor.

In representative agent models, $m$ is often proportional to marginal utility growth. The consumption based capital asset pricing model expresses the pricing kernel as
\[
m_{t+1}=\beta \frac{u'(C_{t+1})}{u'(C_t)}.
\]
Long run risk models, habit formation models, rare disaster models, and intermediary asset pricing models can all be viewed as theories of the pricing kernel \cite{Lucas1978,MehraPrescott1985,CampbellCochrane1999,BansalYaron2004,Barro2006,AdrianEtulaMuir2014}. The measure change from $\mathbb{P}$ to $\mathbb{Q}$ therefore encodes marginal utility and risk compensation.

This perspective also connects to the Hansen Scheinkman decomposition, which separates long term pricing into eigenfunctions, eigenvalues, and martingale components \cite{HansenScheinkman2009}. The long term risk neutral or long forward measure isolates permanent components of valuation. Alvarez and Jermann and Hansen explored how permanent and transitory pricing components affect long horizon asset pricing \cite{AlvarezJermann2005,Hansen2012}. These developments expanded measure changes beyond short dated derivative pricing into macro finance and long horizon valuation.

\section{Benchmark, Real World, and Growth Optimal Pricing}

A different challenge to classical risk neutral pricing comes from benchmark and numeraire portfolio approaches. Long emphasized the numeraire portfolio as a central asset for valuation, and Platen and Heath later developed pricing under the real world measure using the growth optimal portfolio as numeraire \cite{Long1990,PlatenHeath2006}. Under this approach, benchmarked nonnegative portfolios are supermartingales under the physical measure, and fair prices may be computed as real world conditional expectations after normalization by the benchmark. This can allow pricing even when an equivalent risk neutral measure does not exist.

The benchmark approach is historically important because it questions the necessity of an equivalent martingale measure. Classical theory emphasizes the existence of $\mathbb{Q}$ equivalent to $\mathbb{P}$. Benchmark pricing emphasizes the numeraire portfolio and real world dynamics. This is particularly relevant in markets where strict local martingales appear. A discounted asset can be a strict local martingale under a candidate risk neutral measure, creating bubbles and failures of standard put call parity or expectation identities \cite{CoxHobson2005,JarrowProtterShimbo2007,Protter2013}.

Financial bubbles have also been interpreted through measure dependent behavior. Jarrow, Protter, and Shimbo distinguish different forms of bubbles depending on whether prices exceed fundamental values under the relevant pricing measure \cite{JarrowProtterShimbo2007}. The mathematical theory of strict local martingales shows that a process may be a local martingale but not a true martingale, so the expectation representation can fail. This line of research connects asset pricing measures to speculative deviations, nonfundamental values, and the role of the numeraire.

\section{Transaction Costs, Constraints, and Consistent Price Systems}

Classical equivalent martingale measures rely on frictionless trading. With transaction costs, the martingale measure must be replaced or generalized. Jouini and Kallal introduced no arbitrage conditions under transaction costs using consistent price systems \cite{JouiniKallal1995}. Kabanov developed currency market models with proportional transaction costs, where solvency cones replace a single price process \cite{Kabanov1999}. Schachermayer and others refined the fundamental theorem in markets with transaction costs \cite{Schachermayer2004}.

A consistent price system can be thought of as a fictitious frictionless price process lying inside the bid ask spread, together with a martingale measure. It prices trades in a way consistent with the absence of arbitrage under costs. This framework shows that probability measures alone are insufficient when frictions create intervals of prices. The measure must be paired with a shadow price process. Similar issues appear with portfolio constraints, illiquidity, funding costs, collateral, and margin requirements.

The post crisis derivatives literature further complicated the idea of a single risk neutral measure. Collateralized valuation, funding valuation adjustments, credit valuation adjustments, and multiple curves produced pricing formulas depending on collateral accounts, funding policies, counterparty risk, and measure choices \cite{Piterbarg2010,BurgardKjaer2011,BrigoMoriniPallavicini2013}. In these settings, valuation is often nonlinear, and the simple expectation under one equivalent martingale measure may not capture institutional details.

\section{Risk Measures, Ambiguity, and Nonlinear Expectations}

Another major development is the connection between pricing measures and risk measures. Artzner, Delbaen, Eber, and Heath introduced coherent risk measures, which can be represented as worst case expectations over a set of probability measures \cite{ArtznerDelbaenEberHeath1999}. Follmer and Schied developed convex risk measures, extending the representation to penalized families of measures \cite{FollmerSchied2002}. This changed the role of probability measures from a single pricing measure to a set of plausible stress measures.

Robust asset pricing and ambiguity models similarly use multiple priors. Gilboa and Schmeidler's maxmin expected utility and Hansen and Sargent's robust control approach model agents who distrust a single probability law \cite{GilboaSchmeidler1989,HansenSargent2008}. In continuous time, this leads to nonlinear expectations, second order backward stochastic differential equations, and $G$ expectations \cite{Peng2004,SonerTouziZhang2011}. Pricing under ambiguity can therefore be represented as a supremum or infimum over measures, rather than an expectation under one measure.

This development is historically significant because it reverses the classical desire for uniqueness. In robust finance, the multiplicity of measures is not merely a problem caused by market incompleteness. It is a feature reflecting model uncertainty. A price may be conservative, worst case, or ambiguity adjusted. Measure selection becomes part of risk management and regulation.

\section{Information, Filtration Changes, and Enlarged Measures}

Probability measures in asset pricing are also tied to information. A model consists not only of $\mathbb{P}$ or $\mathbb{Q}$, but also of a filtration $(\mathcal{F}_t)_{t\geq 0}$. Insider information, delayed information, partial observation, and news arrivals can change prices by changing conditional distributions. Jacod and Shiryaev's semimartingale theory, Elliott's filtering methods, and credit risk models with progressive enlargement of filtrations all show that asset pricing depends on what is known and when it becomes known \cite{JacodShiryaev2003,ElliottAggounMoore1995,JeanblancYorChesney2009}.

The filtering perspective is especially relevant for modern data driven finance. If investors observe noisy signals about hidden states, then the belief process becomes a state variable. Pricing can be performed under a measure on the enlarged state space containing both fundamentals and beliefs. In equilibrium, prices may depend as much on belief dynamics as on cash flow dynamics. This idea appears in learning models, information based asset pricing, and market microstructure.

Information based asset pricing models, such as those developed by Brody, Hughston, and Macrina, construct price processes from conditional expectations given noisy information processes \cite{BrodyHughstonMacrina2007}. The probability measure determines both future payoffs and the arrival of information about those payoffs. This line anticipates attention based and NLP-based probability transformations, where the flow of textual information affects the perceived distribution of outcomes.

\section{Modern Volatility, Jumps, and Rough Measures}

From the 1990s onward, empirical option pricing focused on richer risk neutral dynamics. Stochastic volatility, jumps, local volatility, affine models, and Levy processes allowed risk neutral distributions to match observed smiles and skews \cite{Heston1993,Bates1996,DuffiePanSingleton2000,ContTankov2004}. These models often distinguish sharply between physical and risk neutral parameters. For example, the variance risk premium is the difference between expected variance under $\mathbb{P}$ and under $\mathbb{Q}$ \cite{CarrWu2009}. The equity index option market reflects a large premium for downside and volatility risk.

Rough volatility is a more recent development. Gatheral, Jaisson, and Rosenbaum documented that volatility behaves as a rough process with low Hurst exponent \cite{GatheralJaissonRosenbaum2018}. Rough Bergomi and related models specify risk neutral dynamics capable of matching short maturity smiles \cite{BayerFrizGatheral2016}. These models do not change the basic martingale measure paradigm, but they show that the risk neutral measure may live on path spaces with highly irregular volatility dynamics.

Machine learning has also entered risk neutral calibration. Neural networks can approximate pricing maps, calibrate volatility surfaces, and learn stochastic discount factors \cite{GuKellyXiu2020,BuehlerGononTeichmannWood2019,ChenPelgerZhu2024}. Recent work through 2025 and 2026 pushes this direction toward transformer based stochastic discount factors, multimodal news embeddings, convex pricing constraints under frictions, and cross asset spillover networks, in which the relevant pricing measure or kernel is represented by a high dimensional learned function of firm, macro, text, frictions, and network information \cite{KellyKuznetsovMalamudXu2025,KorsayeQuainiTrojani2025,WangChengWang2025,AvramovHe2026}. In these approaches, the probability measure may be implicit in simulated paths, generative models, adversarial objectives, learned kernels, or transport maps between the physical and pricing distributions. The classical question remains: how do we transform real world data into pricing relevant expectations while respecting no arbitrage?

\section{Data Driven, Learned, and Information Adjusted Probability Measures}

The most recent period has expanded the role of probability measures without displacing the classical martingale and stochastic discount factor framework. Financial markets now generate large cross sections of firm characteristics, macroeconomic series, option surfaces, order book information, textual news, search data, and social media signals. These data can enter asset pricing in two distinct ways. First, they can help estimate a pricing kernel or stochastic discount factor. Second, they can define an information adjusted forecasting law that reweights future states according to news intensity, attention, sentiment, or disagreement.

\subsection{Learned stochastic discount factors}

The more established modern direction is empirical SDF estimation. Gu, Kelly, and Xiu used machine learning to predict returns and analyze empirical asset pricing anomalies \cite{GuKellyXiu2020}. Chen, Pelger, and Zhu estimated stochastic discount factors with deep neural networks under no arbitrage moment restrictions \cite{ChenPelgerZhu2024}. Recent work extends this idea to transformer based asset pricing models, smart SDFs under convex pricing constraints, and cross asset spillover structures, where the SDF is a learned function of high dimensional conditioning information and market frictions \cite{KellyKuznetsovMalamudXu2025,KorsayeQuainiTrojani2025,AvramovHe2026}. A representative conditional moment restriction is
\begin{equation}
\mathbb{E}\left[m_\theta(X_t,R_{t+1})R_{i,t+1}-1\mid\mathcal{F}_t\right]=0,
\qquad i=1,\ldots,N,
\label{eq:data_sdf_moment}
\end{equation}
where $X_t$ includes firm characteristics, macro variables, option information, or textual embeddings. In this setting, machine learning does not replace probability theory. It estimates the density tilt or pricing kernel that connects $\mathbb{P}$ to pricing restrictions.

\subsection{Text, news, and attention as conditioning information}

Text based finance has a long history, including media pessimism, financial dictionaries, tone measures, and return prediction with text \cite{Tetlock2007,LoughranMcDonald2011,JegadeeshWu2013,KeKellyXiu2019}. Transformer architectures and large language models have made textual signals more flexible and context aware \cite{Vaswani2017,DevlinChangLeeToutanova2019,LopezLiraTang2023}. Recent SDF work using news embeddings illustrates the same trend: unstructured text may become part of the information set used to estimate a pricing kernel \cite{WangChengWang2025}. In measure language, text changes the conditional distribution
\begin{equation}
\mathbb{P}(R_{t+1}\in A\mid\mathcal{F}_t)
\quad \hbox{to} \quad
\mathbb{P}(R_{t+1}\in A\mid\mathcal{F}_t\vee\mathcal{N}_t),
\label{eq:text_conditioning}
\end{equation}
where $\mathcal{N}_t$ denotes the sigma field generated by news and textual signals. This is a filtration enlargement rather than necessarily a new arbitrage pricing measure.

\subsection{Information adjusted forecast tilts from text and attention}

Beyond conditioning on text, some applications translate information variables into explicit probability tilts. This representation is natural when observed news flow is not a neutral sample of economically relevant states. Media coverage, analyst attention, and social amplification may overweight some issuers or narratives; a forecasting law can then be adjusted by a positive Radon--Nikodym density. Cao and Geman's hype adjusted probability measure fits into this information-tilting view \cite{CaoGeman2025HAP, cao2025hypeindex, cao2025hlppl}. It reweights NLP-based return and volatility forecasts using attention, bias correction, and sentiment memory, while no arbitrage valuation remains governed by an SDF or martingale measure. The later papers extend the hype-adjusted probability measure into the hype index and applications on identifying and quantifying financial bubbles and their reversals.

A general mathematical version is as follows. Let $(\Omega,\mathcal{F},\mathbb{P})$ be a baseline forecasting space and let $A_t$ be an $\mathcal{F}_t$ measurable attention or news variable. For an asset $i$, define news and market weights by
\begin{equation}
n_{i,t}=\frac{N_{i,t}}{\sum_{j=1}^m N_{j,t}},
\qquad
c_{i,t}=\frac{C_{i,t}}{\sum_{j=1}^m C_{j,t}},
\qquad
b_{i,t}=n_{i,t}-c_{i,t},
\label{eq:balanced_attention_imbalance}
\end{equation}
where $N_{i,t}$ is the number of relevant news items and $C_{i,t}$ is a market capitalization or economic weight. A positive $b_{i,t}$ indicates that the asset is over represented in the news relative to size; a negative value indicates under representation. Sentiment memory can be represented by
\begin{equation}
\bar S_{i,t}=\sum_{k=0}^{K}\rho^k w_{i,t-k}S_{i,t-k},
\qquad 0<\rho<1,
\label{eq:balanced_memory_sentiment}
\end{equation}
where $S_{i,t-k}$ is a sentiment score and $w_{i,t-k}$ can incorporate source reliability or sector weights.

The adjusted forecasting measure $\mathbb{P}^{a}$ is defined by a positive normalized density:
\begin{equation}
Z_T^a>0,
\qquad
\mathbb{E}^{\mathbb{P}}[Z_T^a]=1,
\qquad
\frac{d\mathbb{P}^{a}}{d\mathbb{P}}=Z_T^a.
\label{eq:balanced_info_density}
\end{equation}
A convenient exponential tilting rule is
\begin{equation}
Z_T^a(\omega)=
\frac{\exp\{\theta_1 A_T(\omega)+\theta_2 B_T(\omega)+\theta_3 \bar S_T(\omega)\}}
{\mathbb{E}^{\mathbb{P}}\left[\exp\{\theta_1 A_T+\theta_2 B_T+\theta_3 \bar S_T\}\right]},
\label{eq:balanced_exponential_tilt}
\end{equation}
where $A_T$ is an attention factor, $B_T$ is a bias correction factor, and $\bar S_T$ is a memory adjusted sentiment factor. Forecasts are computed by Bayes' rule:
\begin{equation}
\mathbb{E}^{\mathbb{P}^{a}}[X\mid\mathcal F_t]
=\frac{1}{Z_t^a}\mathbb{E}^{\mathbb{P}}[Z_T^aX\mid\mathcal F_t],
\qquad
Z_t^a=\mathbb{E}^{\mathbb{P}}[Z_T^a\mid\mathcal F_t].
\label{eq:balanced_info_bayes}
\end{equation}
For forecasting, $X$ may be a return, a volatility measure, or an event indicator. For derivative pricing, however, the relevant object remains a risk neutral measure or an SDF satisfying no arbitrage restrictions.

The distinction is important. A risk neutral measure $\mathbb{Q}$ must satisfy
\begin{equation}
B_t^{-1}S_t=\mathbb{E}^{\mathbb{Q}}[B_T^{-1}S_T\mid\mathcal{F}_t],
\label{eq:q_martingale_info_section}
\end{equation}
whereas an information adjusted forecasting measure generally need not. If both are used, the inferred pricing kernel changes according to
\begin{equation}
\frac{d\mathbb{Q}}{d\mathbb{P}^{a}}
=\frac{d\mathbb{Q}/d\mathbb{P}}{d\mathbb{P}^{a}/d\mathbb{P}}
=\frac{Z_T^{\mathbb{Q}}}{Z_T^{a}}.
\label{eq:balanced_q_vs_pa}
\end{equation}
Thus an information-adjusted forecasting tilt is best interpreted as one way to model the information content of $\mathbb{P}$, or as an input to a learned SDF, not as a substitute for martingale pricing.

This line connects naturally to behavioral and attention based asset pricing. Barber and Odean studied attention grabbing stocks, Da, Engelberg, and Gao used search volume as a direct attention measure, and Bordalo, Gennaioli, and Shleifer modeled diagnostic expectations in which representative states are overweighted \cite{BarberOdean2008,DaEngelbergGao2011,BordaloGennaioliShleifer2018}. The measure theoretic contribution of attention adjusted forecasting is to write these effects as explicit density tilts. The main open issues are measurability, look ahead bias, stability, and economic interpretation. A regularized version can be written as
\begin{equation}
Z^{a,*}=\argmin_{Z>0,\,\mathbb{E}^{\mathbb{P}}[Z]=1}
\left\{\mathcal L(Z;\hbox{forecast target})+\eta\,\mathbb{E}^{\mathbb{P}}[Z\log Z]\right\},
\label{eq:balanced_regularized_info}
\end{equation}
which links information adjusted probability tilts to entropy penalization, robust control, and learned stochastic discount factors.

\section{Key Formula Compendium by Measure Type}

This section gathers the formulas most frequently used in the historical literature. It is included to make explicit the mathematical assumptions behind the narrative review.

\subsection{Capital asset pricing and linear pricing kernels}

In the CAPM, expected excess returns satisfy
\begin{equation}
\mathbb{E}[R_i]-R_f=\beta_i\left(\mathbb{E}[R_M]-R_f\right),
\qquad
\beta_i=\frac{\Cov(R_i,R_M)}{\Var(R_M)}.
\label{eq:capm}
\end{equation}
A corresponding linear stochastic discount factor can be written as
\begin{equation}
m=a-bR_M,
\label{eq:linearsdf}
\end{equation}
with constants $a,b$ chosen so that $\mathbb{E}[mR_i]=1$. The formula shows why mean variance theory is an early form of probability tilting: high beta assets have payoffs in states where the pricing kernel is low, so they require higher expected returns.

\subsection{Black Scholes Merton pricing}

For a non dividend paying stock,
\begin{equation}
dS_t=\mu S_tdt+\sigma S_tdW_t^{\mathbb P},
\qquad
S_0>0.
\label{eq:bsmphysical}
\end{equation}
Under the risk neutral measure,
\begin{equation}
dS_t=rS_tdt+\sigma S_tdW_t^{\mathbb Q}.
\label{eq:bsmriskneutral}
\end{equation}
For a European call $H=(S_T-K)^+$, the price is
\begin{equation}
C_t=S_t\Phi(d_1)-Ke^{-r(T-t)}\Phi(d_2),
\label{eq:bsmcall}
\end{equation}
where
\begin{equation}
d_1=\frac{\log(S_t/K)+(r+\frac12\sigma^2)(T-t)}{\sigma\sqrt{T-t}},
\qquad
d_2=d_1-\sigma\sqrt{T-t}.
\label{eq:d1d2}
\end{equation}
The remarkable feature is that the physical drift $\mu$ does not enter the option price. It is absorbed into the change from $\mathbb P$ to $\mathbb Q$.

\subsection{Change of numeraire}

Let $N_t>0$ be a traded numeraire and let $\mathbb Q^B$ be the money market measure associated with $B_t$. The $N$ numeraire measure is defined by
\begin{equation}
\frac{d\mathbb Q^N}{d\mathbb Q^B}\bigg|_{\mathcal F_T}
=\frac{B_0N_T}{N_0B_T}.
\label{eq:numerairedensity}
\end{equation}
If $V_t$ is the price of payoff $H$, then
\begin{equation}
\frac{V_t}{N_t}=\mathbb E^{\mathbb Q^N}\left[\frac{H}{N_T}\mid\mathcal F_t\right].
\label{eq:numerairepricing}
\end{equation}
Choosing $N_t=P(t,T)$, a zero coupon bond maturing at $T$, gives the $T$ forward measure. Under that measure, many forward prices and swap rates become martingales.

\subsection{Term structure drift restrictions}

In the Heath Jarrow Morton framework, the instantaneous forward rate satisfies
\begin{equation}
df(t,T)=\alpha(t,T)dt+\sigma(t,T)dW_t^{\mathbb Q}.
\label{eq:hjm}
\end{equation}
No arbitrage under the money market numeraire imposes the drift restriction
\begin{equation}
\alpha(t,T)=\sigma(t,T)\int_t^T \sigma(t,u)du
\label{eq:hjmdrift}
\end{equation}
in the one factor case, with the natural inner product generalization in the multifactor case. This is a canonical example of a probability measure determining the drift of an entire curve.

\subsection{Local volatility and implied risk neutral density}

If call prices $C(K,T)$ are sufficiently smooth, the Breeden Litzenberger identity extracts the risk neutral density:
\begin{equation}
f_{\mathbb Q}(K,T)=e^{rT}\frac{\partial^2 C(K,T)}{\partial K^2}.
\label{eq:breedenlitzenberger}
\end{equation}
The Dupire local volatility formula can be written as
\begin{equation}
\sigma_{\mathrm{loc}}^2(K,T)
=\frac{\partial_T C(K,T)+rK\partial_K C(K,T)}{\frac12K^2\partial_{KK}C(K,T)}.
\label{eq:dupire}
\end{equation}
These formulas reveal that option markets contain an implied probability distribution, although the distribution is risk neutral rather than physical.

\subsection{Incomplete market selection rules}

When $\mathcal M_e$ has many elements, a measure can be selected by minimizing a criterion. The minimal entropy martingale measure solves
\begin{equation}
\mathbb Q^{\mathrm{ME}}
=\argmin_{\mathbb Q\in\mathcal M_e}
\mathbb E^{\mathbb P}\left[\frac{d\mathbb Q}{d\mathbb P}\log\frac{d\mathbb Q}{d\mathbb P}\right].
\label{eq:minentropy}
\end{equation}
The variance optimal martingale measure minimizes
\begin{equation}
\mathbb E^{\mathbb P}\left[\left(\frac{d\mathbb Q}{d\mathbb P}\right)^2\right]
\label{eq:varianceoptimal}
\end{equation}
over signed or equivalent martingale measures, depending on the formulation. Utility indifference pricing defines a price $p$ by the equation
\begin{equation}
\sup_{\theta}\mathbb E\left[U\left(X_T^{x,\theta}\right)\right]
=\sup_{\theta}\mathbb E\left[U\left(X_T^{x-p,\theta}+H\right)\right].
\label{eq:utilityindifference}
\end{equation}
These equations illustrate that incomplete market pricing turns measure choice into an optimization problem.

\subsection{Coherent and convex risk measures}

A coherent risk measure has the dual representation
\begin{equation}
\rho(X)=\sup_{\mathbb Q\in\mathcal Q}\mathbb E^{\mathbb Q}[-X],
\label{eq:coherentrisk}
\end{equation}
where $\mathcal Q$ is a set of probability measures. A convex risk measure has the more general representation
\begin{equation}
\rho(X)=\sup_{\mathbb Q\ll\mathbb P}\left\{\mathbb E^{\mathbb Q}[-X]-\alpha(\mathbb Q)\right\},
\label{eq:convexrisk}
\end{equation}
where $\alpha$ is a penalty function. Pricing under ambiguity therefore uses a family of measures rather than a single transformed measure.

\subsection{Benchmark and real world pricing}

Let $S^*_t$ denote the growth optimal portfolio or numeraire portfolio. In benchmark pricing, a payoff $H$ can be priced by
\begin{equation}
V_t=S^*_t\,\mathbb E^{\mathbb P}\left[\frac{H}{S^*_T}\mid\mathcal F_t\right],
\label{eq:benchmarkpricing}
\end{equation}
provided the benchmarked price process is a martingale or the appropriate supermartingale valuation is used. This formula emphasizes that valuation can sometimes be performed under the physical measure when the numeraire is chosen appropriately.

\subsection{Machine learning and learned stochastic discount factors}

Recent empirical asset pricing often parameterizes the stochastic discount factor by a neural network or other flexible function $m_\theta(X_t,R_{t+1})$ and estimates it from moment restrictions such as
\begin{equation}
\mathbb E\left[m_\theta R_{i,t+1}-1\right]=0,
\qquad i=1,\ldots,d.
\label{eq:learnedsdfmoment}
\end{equation}
A regularized empirical criterion is
\begin{equation}
\min_{\theta}\left\|\frac{1}{T}\sum_{t=1}^T g_t\left(m_\theta R_{t+1}-\mathbf 1\right)\right\|^2+\lambda\,\mathcal R(\theta),
\label{eq:learnedsdfobjective}
\end{equation}
where $g_t$ are instruments and $\mathcal R$ is a regularizer. In this language, modern machine learning does not abandon probability measures. It estimates high dimensional pricing kernels or implicit measure changes.

\section{A Historical Taxonomy}

Table \ref{tab:taxonomy} summarizes the main probability measures and measure related objects in asset pricing.

\begin{table}[!htbp]
\centering
\small
\renewcommand{\arraystretch}{1.15}
\begin{tabular}{p{0.25\textwidth}p{0.32\textwidth}p{0.33\textwidth}}
\toprule
Object & Main idea & Historical role \\
\midrule
Physical measure $\mathbb{P}$ & Statistical law of returns & Forecasting and econometrics \\
State price measure & Normalized Arrow Debreu prices & Static contingent claim valuation \\
Risk neutral measure $\mathbb{Q}$ & Discounted prices are martingales & Derivative pricing \\
Equivalent martingale measure & $\mathbb{Q}\sim\mathbb{P}$ and no arbitrage & Fundamental theorem \\
Forward measure & Bond numeraire measure & Interest rate derivatives \\
Pricing kernel & $p(X)=\mathbb{E}^{\mathbb{P}}[mX]$ & Equilibrium and empirical asset pricing \\
Minimal martingale measure & Preserve orthogonal risk & Incomplete market hedging \\
Minimal entropy measure & Minimize relative entropy & Incomplete market selection \\
Utility indifference measure & Marginal utility based pricing & Preference based pricing \\
Benchmark measure & Growth optimal numeraire & Real world pricing \\
Robust measure set & Worst case or penalized measures & Ambiguity and risk management \\
Information/attention adjusted measure & Text, news, and attention reweighting & Forecasting and learned-SDF conditioning \\
\bottomrule
\end{tabular}
\caption{A taxonomy of probability measures and related valuation objects.}
\label{tab:taxonomy}
\end{table}

The table highlights a recurring pattern. Each new market feature creates a new measure concept. Dynamic hedging creates the risk neutral measure. Interest rate numeraires create forward measures. Incomplete markets create selection criteria. Equilibrium creates stochastic discount factors. Transaction costs create consistent price systems. Ambiguity creates sets of measures. Textual information creates attention adjusted measures. The historical development is therefore cumulative rather than substitutive.

\section{Open Problems and Directions Toward 2026}

Several open directions remain important in 2026. First, the relation between physical and risk neutral measures is still central to empirical finance. Estimating risk premia requires comparing $\mathbb{P}$ and $\mathbb{Q}$, but both are difficult to identify. Option prices provide rich information about $\mathbb{Q}$, while historical data, surveys, macro variables, and machine learning models inform $\mathbb{P}$. The bridge between them remains a major research frontier.

Second, incomplete market measure selection is increasingly relevant. Climate risk, geopolitical risk, cyber risk, AI risk, and prediction market event risk are often only partially hedgeable. A unique risk neutral measure is unrealistic. Pricing must combine observed traded claims, subjective beliefs, robust sets, and equilibrium restrictions.

Third, data-driven and information-adjusted probability measures need stronger theoretical foundations. NLP and alternative data can improve forecasting, but a probability reweighting rule should be measurable with respect to available information, interpretable, stable, and economically justified. In this respect, Cao and Geman's hype adjusted probability measure is a natural recent example of the broader attention-based literature: it expresses news coverage, sentiment, and representativeness as a normalized probability tilt rather than as a stand-alone pricing law \cite{CaoGeman2025HAP}. Future work may connect attention adjusted densities to equilibrium disagreement, market microstructure, learned SDFs, and option implied risk premia.

Fourth, no arbitrage constraints for machine learning models remain crucial. A neural model that predicts prices must avoid static and dynamic arbitrage. The newest AI asset pricing work emphasizes transformer SDFs, adversarial moment restrictions, textual embeddings, and cross asset spillover structures, but these methods still need transparent links between statistical prediction, pricing kernels, and admissible probability measures \cite{ChenPelgerZhu2024,KellyKuznetsovMalamudXu2025,WangChengWang2025,AvramovHe2026}. One promising direction is to learn stochastic discount factors, martingale measures, or transport maps between physical and pricing distributions under explicit constraints. Another is to combine generative models with calibration to liquid option surfaces and historical path features.

Fifth, the rise of prediction markets and event contracts may expand the traded state space. When new contracts are listed on inflation, elections, macro announcements, sports outcomes, or climate events, they create partial state prices for previously untraded risks. This may allow researchers to infer event specific pricing measures and compare them with survey beliefs and statistical probabilities. Such markets provide a natural laboratory for studying how probability measures are formed from prices.

\section{Conclusion}

The concept of a probability measure in asset pricing has undergone a remarkable transformation. In early mathematical finance, probability described random price movements. In Arrow Debreu theory, normalized state prices became probabilities. In Black Scholes Merton, the risk neutral measure removed expected returns from derivative pricing. In martingale theory, equivalent martingale measures became the mathematical expression of no arbitrage. In term structure modeling, the numeraire determined the measure. In incomplete markets, measure selection became a matter of hedging, entropy, utility, or calibration. In equilibrium asset pricing, the stochastic discount factor tilted physical probabilities by marginal utility. In robust finance, sets of measures represented ambiguity. In modern data driven finance, information and attention variables can define new probability transformations.

The broad lesson is that there is no single probability measure for asset pricing. There are physical measures for statistical dynamics, pricing measures for valuation, forward measures for numeraires, kernels for preferences, robust measure sets for ambiguity, and information adjusted measures for forecasting. Recent information adjusted measures are useful because they show how new information sources can be translated into probability tilts, but the core historical structure remains the classical sequence from state prices to martingale measures, numeraire measures, and stochastic discount factors. As financial markets become more data rich and information intensive, the future of asset pricing will likely involve not only better models under a fixed probability measure, but better theories of how probability measures themselves are constructed, transformed, selected, and interpreted.

\end{document}